\documentclass[aps,pra,twocolumn,groupedaddress,floatfix]{revtex4}
\usepackage{graphicx}

\begin{document}
\title{On the Prospects for Laser Cooling of TlF}
\date{\today}
\author{L.R. Hunter}
\affiliation{Physics Department, Amherst College, Amherst, Massachusetts 01002, USA}
\author{S.K. Peck}
\affiliation{Physics Department, Amherst College, Amherst, Massachusetts 01002, USA}
\author{A.S. Greenspon}
\affiliation{Physics Department, Amherst College, Amherst, Massachusetts 01002, USA}
\author{S. Saad Alam}
\affiliation{Physics Department, Amherst College, Amherst, Massachusetts 01002, USA}
\author{D. DeMille}
\affiliation{Department of Physics, P.O. Box 208120, Yale University, New Haven, Connecticut 06520, USA}
\begin{abstract}
We  measure the upper state lifetime and two ratios of vibrational branching fractions \(f_{v'v}\) on the \(B^{3}\Pi_{1}(v') - X^{1}\Sigma^{+}(v)\) transition of TlF.  We find the B state lifetime to be 99(9) ns. We also determine that the off-diagonal vibrational decays are highly suppressed: \(f_{01}/f_{00} < 2 \times 10^{-4}\) and \(f_{02}/f_{00} = 1.10(6)\)\% , in excellent agreement with their predicted values of \(f_{01}/f_{00} < 8 \times 10^{-4}\) and \(f_{02}/f_{00} = 1.0(2)\)\% based on Franck-Condon factors calculated using Morse and RKR potentials.  The implications of these results for the possible laser cooling of TlF and fundamental symmetries experiments are discussed.
\end{abstract}


\maketitle

The laser cooling of molecules presents a daunting challenge with potentially rich rewards.   Transverse cooling and collimation of a cold molecular beam can be accomplished by scattering a few hundred photons from each molecule, while over 10,000 photons must be scattered to bring a typical molecule in the beam to rest.  It is difficult to find molecular cycling transitions that will allow so many absorption and decay cycles without significant loss to the myriad of rotational and vibrational states present in most molecular systems.  In addition, successful laser cooling requires that the excited state be relatively short lived, such that it can complete many absorption/emission cycles before leaving the laser interaction region.  Despite these challenges, transverse cooling has recently been achieved in strontium monofluoride \cite{ref01,ref02}.

In the present paper we explore the possibility of laser cooling thallium monofluoride (TlF).  High precision searches for the Schiff moment and the proton electric dipole moment (EDM) have been carried out in beams of TlF \cite{ref03,ref04}. These experiments are tests of both parity (P) and time-reversal (T) symmetries.  TlF exhibits a large enhancement in its sensitivity to such violations due to the large internal electric field of the molecule and thallium's large atomic number \(Z\! =\! 81\) \cite{ref05}.  It also displays a remarkable insensitivity to systematic effects associated with external magnetic fields.  The TlF experiments were largely limited by the relatively broad line widths associated with the molecules' rapid transit time through the apparatus, and the modest thermal populations of the state of interest.  Using cryogenic beams and laser cooling, it may be possible to overcome these limitations and TlF might again emerge as an interesting candidate for measuring symmetry violations in the nucleus.

The transition \(X^{1}\Sigma^{+} (v\! =\! 0, J^{P}\! =\! 1^{-}) - B^{3}\Pi_{1} (v'\! =\! 0, J^{\prime P}\! =\! 1^{+})\) (where \(J^{P}\) denotes the rotational angular momentum and parity) of TlF is an interesting candidate for a cycling transition.  As we argue here, it appears that this transition should be highly closed to other (unwanted) electronic, vibrational, and rotational decay paths. We begin with a discussion of electronic decay paths. The only other electronic decay transition from the \(B^{3}\Pi_{1}\) state is to the \(A^{3}\Pi_{0^+}\) state (and its as-yet unobserved \({}^{3}\Pi_{0^-}\) partner).  The branching fraction for this transition should be very small, according to the following logic.  First, the transition frequency \(\omega_{BA}\) for the \textit{B-A} transition is much smaller than that for the \textit{B-X} transition (\(\omega_{BX}\)): \(\omega_{BX}/\omega_{BA} \approx 22\). Moreover, the electric dipole (E1) matrix element \(d_{BA}\)  should be several times smaller than \(d_{BX}\). This can be seen as follows.  Nominally, both the \textit{B-X} and \textit{B-A} E1 transitions are forbidden: the former requires a change in the total spin S and the latter a change in the projection \(\Sigma\) of \textit{S} along the internuclear axis.  Both transitions acquire a non-zero matrix element due to off-diagonal mixing with other electronic states, via the spin-orbit (SO) interaction.  For the \textit{B-X} transition, this occurs primarily due to mixing of the nearby \(6s\sigma 6p\pi\; C^{1}\Pi_{1}\) state into the \(6s\sigma 6p\pi\; B^{3}\Pi_{1}\) state; the admixed \(6s\sigma 6p\pi\; C^{1}\Pi_{1} - (6s\sigma)^{2} X^{1} \Sigma^{+}\) Rydberg transition is very strong.  However, due to selection rules for SO mixing (\(\Delta S \! =\! \pm 1, 0\); \(\Delta\Sigma \! =\! \pm 1, 0\); \(\Delta\Lambda \! =\! \pm 1,0\) , where \(\Lambda\) is the projection of orbital angular momentum along the internuclear axis; and \(\Delta\Omega \! =\!  0\), where \(\Omega \! =\!  \Lambda + \Sigma\)) \cite{ref06}, it can be verified that the \textit{B-A} transition requires SO mixing with another orbital configuration (or a second-order mixing if only the lowest-lying configurations are considered).  The nearest relevant level is likely the as yet unobserved \(6s\sigma 6p\sigma\; {}^{3}\Sigma^{+}\) state, which can mix into both the \(B\) state and the \({}^{3}\Pi_{0^-}\) state; however, this admixture leads only to a weak perpendicular-band \cite{ref07} (\(6p\pi - 6p\sigma\)) transition.  Taking into account all possibilities, we crudely estimate \(d_{BX}/d_{BA} \approx 3\).  Finally, since the decay rate \(\Gamma \propto d^{2}\omega^{3}\), we estimate that \(\Gamma_{BX}/\Gamma_{BA} \sim 10^{5}\). As such, unwanted \textit{B-A} decays are unlikely to significantly limit the cooling process.

Selection rules limit the rotational states of \(X^{1}\Sigma^{+}\) accessible to the decay from the \(B^{3}\Pi_{1} (v' \! =\! 0, J^{\prime\, P} \! =\! 1^{+})\) state.  In the absence of hyperfine structure (HFS), this state can decay \textit{only} to \(X^{1}\Sigma^{+} (J^{P} \! =\! 1^{-})\).  Off-diagonal mixing of different rotational levels due to HFS can in principle allow unwanted rotational transitions.  In the \(X^{1}\Sigma^{+}\) state the HFS is very small ($\sim$kHz)\cite{ref03} and such mixing is negligible.  In the \(B^{3}\Pi_{1}\) state, the HFS interaction can be estimated from a simple scaling law (HFS is roughly proportional to \textit{Z}) combined with the observed HFS in the isoelectronic species AlF \cite{ref08}.  Based on this, HFS interaction with the Tl nucleus (\(I_{\mathrm{Tl}} \! =\! \frac{1}{2}\) for both \({}^{205}\mathrm{Tl}\) and \({}^{203}\mathrm{Tl}\)) can be substantial, with estimated \(B^{3}\Pi_{1} (v' \! =\!  0, J^{\prime\, P} \! =\!  1^{+})\) state HFS interaction strength of order \(A_{\mathrm{Tl}} \sim 1\)~GHz, while HFS associated with the \({}^{19}\mathrm{F}\) nucleus (\(I_{\mathrm{F}} \! =\!  \frac{1}{2}\)) should be much smaller (\(A_{\mathrm{F}} \approx 10\)~MHz).  Hence we define the intermediate angular momentum \(F_{1} \! =\!  J + I_{\mathrm{Tl}}\) and the total angular momentum \(F \! =\!  F_{1} + I_{\mathrm{F}}\).  The \(B^{3}\Pi_{1} (v' \! =\!  0, J^{\prime\, P} \! =\!  1^{+})\) state should split into well-resolved levels with \(F'_{1} = \frac{1}{2}~ \mbox{ and } \frac{3}{2}\), each of which will consist of a pair of closely-spaced levels with \(F' = F'_{1} \pm \frac{1}{2}\).  Cycling on the \(X^{1}\Sigma^{+} (J^{P} \! =\!  1^{-}, F_{1},F) - B^{3}\Pi_{1} (v' \! =\!  0, J^{\prime\, P} \! =\!  1^{+}, F'_{1} \! =\!  \frac{1}{2}, F')\) manifold of nearly-degenerate transitions should have minimal loss to other \(X^{1}\Sigma^{+} (J^{\prime\prime})\) levels, since these \textit{B} state sublevels mix with levels of higher \(J'\) only due to the weak HFS interaction of \({}^{19}\mathrm{F}\).  In this case, the branching fraction \(f_{J'J''}\) to unwanted higher rotational states can be estimated as \(f_{J'J''} \sim (A_{\mathrm{F}}/B_e^B)^2 < 10^{-6}\), where \(B_e^B\) is the rotational constant in the \(B^{3}\Pi_{1}\) state.  Thus, using this particular HFS transition should enable cooling with little loss.

Because no selection rules limit the various possible vibrational transitions, branching to unwanted vibrational levels is typically the most challenging problem in the laser cooling of a molecule \cite{ref09}.  We have numerically calculated Franck-Condon factors (FCF) for the low lying vibrational transitions under various assumptions and approximations, in an attempt to quantify the uncertainties in their values.  In all cases, precise spectroscopic data, suitable for determining both the \(B^{3}\Pi_{1}\) and \(X^{1}\Sigma^{+}\) state potentials, is taken from Ref.~\cite{ref10}.  In one approach, we used the full set of tabulated Dunham coefficients for both states, using standard routines for determining RKR potentials and their associated FCFs \cite{ref11}.  However, we believe there are two significant ambiguities in this approach.  First, the difference in effective rotational constants between \textit{e} and \textit{f}-parity levels in the \(B^{3}\Pi_{1}\) state, due to \(\Omega\)-doubling, has a non-negligible effect on the FCFs at the precision of interest here; it is not clear whether this difference is physically meaningful in terms of its effect on the molecular potentials.  Second, the data in Ref.~\cite{ref10} includes the Dunham coefficient \(Y_{30}\) (second vibrationally anharmonic term) only for the \(B^{3}\Pi_{1}\) state, while the Dunham expansion terminates at \(Y_{20}\) for the \(X^{1}\Sigma^{+}\) state; it is unclear whether this inherent asymmetry in treatment of the potentials can lead to errors in the FCFs, particularly for higher vibrational levels of the \textit{X} state.  Hence, we evaluated FCFs for RKR potentials with rotational Dunham coefficiencts of both \textit{e} and \textit{f}-parity levels, and both with and without inclusion of the \(Y_{30}\) term for the \textit{B} state.  Finally, in order to quantify possible errors in the FCFs due to the stated uncertainties in the spectroscopic data, we evaluated the FCFs using Morse potentials and the associated analytic form of the vibrational wavefuctions \cite{ref12}. Here we used the Dunham coefficients \(Y_{10}\) and \(Y_{20}\) for the harmonic and anharmonic vibrational terms of the Morse potential; for the \textit{X} state we used the Dunham coefficient \(Y_{01}\) as the effective rotational constant to determine the internuclear separation \(r_{e} = 2.084438\)~\AA, while for the \textit{B} state we used \(r'_{e} = 2.0740(5)\) ~\AA ~as derived in Ref.~\cite{ref10} from a combined potential fit to both \textit{e}- and \textit{f}-parity levels.  Within this model we calculated FCFs with all input parameters varied within their stated ranges of uncertainty.

We quote a total uncertainty range for our calculated FCFs that incorporates all values obtained from these various calculations.  The resulting Franck-Condon factors are shown in Table~\ref{tab01}.
\begin{table*}
\begin{tabular}{|c||c|c|c|c|c|c|c|}\hline
\(v'\backslash v\) & 0 & 1 & 2 & 3 & 4 & 5 & \(>6\)\\ \hline\hline
0 & 0.989(2) & \(< 0.0008\) & 0.011(2) & \(< 0.0003\) & \(< 0.0002\) & \(< 0.0003\) & \(< 0.0002\)\\ \hline
1 & \(< 0.0004\) & 0.949(6) & 0.016(8) & 0.031(7) & \(< 0.006\) & \(< 0.0011\) & \(< 0.002\)\\ \hline
2 & 0.010(2) & 0.03(2) & 0.77(6) & 0.09(5) & 0.08(2) & 0.03(2) & \(< 0.006\)\\ \hline
\end{tabular}
\caption{\small Calculated Franck-Condon Factors for the \(X^{1}\Sigma^{+} - B^{3}\Pi_{1}\) transition of TlF.  The primed (unprimed) vibrational quantum numbers refer to the excited (ground) electronic level.  As described in the text, the central values and the uncertainties have been chosen to accommodate the variations in the predictions from the different model potentials and the uncertainties in the spectroscopic parameters used as input to the models.\label{tab01}}
\end{table*}
The FCF matrix is highly diagonal, so it appears that the \(X(v \! =\!  0)\) to \(B(v' \! =\!  0)\) transition could provide a good cycling transition, with leakage to other vibrational levels broadly similar to that of the demonstrated case of SrF \cite{ref01,ref02}.  However, the uncertainties in the FCFs are significant at the level needed to evaluate exactly how many, and which, vibrational repumping lasers would be needed in order to scatter enough photons for laser cooling and/or slowing of TlF.  In addition, the \(B^{3}\Pi_{1}\) potential of TlF, while mostly arising from an ionic bond between \(\mathrm{Tl}^{+}\) and \(\mathrm{F}^{-}\), is substantially modified by a curve crossing with a covalent-bonding potential \cite{ref13}.  Hence, it seems conceivable that the \textit{X-B} transition dipole moment could change significantly with the internuclear distance; in this case the FCFs alone would not be sufficient to determine the vibrational branching fractions.   
Given all the possible sources of error, we chose to measure two of the vibrational branching ratios to test the accuracy of our calculations.  

Our measurements were made using a molecular beam that originates from a stainless steel oven filled with TlF.  The oven is heated to temperatures in the range 415--460 C, vaporizing the TlF which escapes the oven through a hole in the chamber.  A set of four hollow ceramic tubes (25 mm long, 2.4 mm id and 3.3 mm od) fill this opening.  These tubes are kept warmer than the oven and serve to pre-collimate the beam.  The beam is further collimated by a 6 mm high by 10 mm wide aperture located about 30 cm from the oven.   The interaction region where the TlF beam intersects our laser beam is located about 6.5 cm downstream from this aperture.  This entire beam assembly is contained in a cylindrical stainless steel vacuum chamber which is maintained at a pressure of about \(10^{-6}\) Torr by diffusion pumps.

In order to excite the \(X(v \! =\!  0)\) to \(B(v' \! =\!  0)\) transition we need to generate ultraviolet (uv) light at 271.7 nm.  We use a tripled YAG laser (Quantel Model 770B) at 355 nm to pump a dye laser (Quantel Model TDL60) with Coumarin 540A dye.  The output of this laser is tunable near 543.4 nm.  The beam is directed to a BBO crystal where it is frequency doubled to produce the desired uv wavelength.  The residual green light emerging from the BBO crystal is directed to a wavemeter (New Focus Model 7711) where its wavelength is monitored. The uv laser beam is incident upon the molecular beam at a right angle and their overlap volume defines the region of the molecular excitation.   To minimize scattering of the laser light, the linearly polarized laser beam enters and exits the vacuum chamber through quartz Brewster windows, mounted about 38 cm from the molecular beam.

Detectors are placed above and below this interaction region, on a line perpendicular to both the laser and molecular beams and parallel to the laser polarization.  Fluorescence from the B state passes out of the vacuum chamber through a window, is collimated by a lens, passes through an interference filter and is focused on an adjustable diameter aperture that serves as a spatial filter.  The fluorescence which passes through the aperture is detected by a uv sensitive phototube.  All optics are fused silica to allow transmission of the uv light.  Following preamplification, the single photon signals are counted by a photon counter (Stanford Research model SR400).  The device counts the number of photons collected by each of our two detectors and relays this information to a computer for storage.

\begin{figure}
\centering
\includegraphics[width=3.25in]{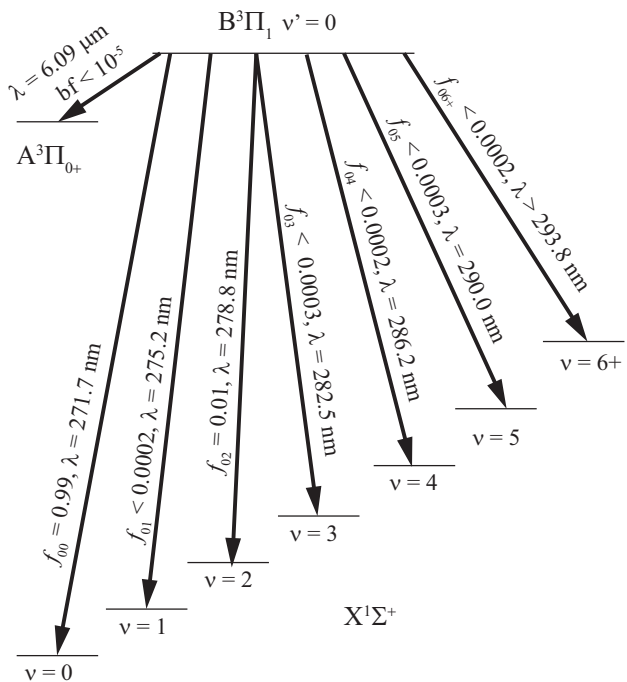}
\caption{\small The possible decay paths of the \(v' \! =\!  0\) level of the \(B^{3}\Pi_{1}\) state.  The values we have deduced for the branching fractions \(f_{0v}\) to the various vibrational levels of the \(X^{1}\Sigma^{+}\) state are shown.  Our estimated upper bound on the branching fraction (bf) to the  \(A^{3}\Pi_{0^+}\) state is also shown.}
\label{fig01}
\end{figure}

\begin{table}
\begin{tabular}{|c|c|c|}\hline
Branching ratio & Theory & Experiment\\ \hline  
\(f_{01}/f_{00}\) & \(< 0.0008\) & \(< 0.0002\)\\ \hline
\(f_{02}/f_{00}\) & 0.010(2) & 0.0110(6)\\ \hline
\end{tabular}
\caption{\small A comparison of our measured and predicted ratios of branching fractions from the \(B^{3}\Pi_{1}, v' \! =\!  0\) state to the various vibrational states \textit{v} of the \(X^{1}\Sigma^{+}\) level in TlF.  The branching fraction is taken as proportional to the Franck-Condon factor times the cube of the decay energy for that branch.\label{tab02}}
\end{table}
We use the lower detector to constantly monitor the strong fluorescence signal at 271.7 nm.  This provides a useful relative calibration of the fluorescence intensity and allows us to remove fluctuations in laser intensity and frequency and molecular beam intensity from our measurements.  To measure branching ratios, we alternate between three different interference filters in the upper optics assembly.  One of these is chosen to transmit the strong \(v' \! =\!  0\) to \(v \! =\!  0\) fluorescence near 271.7 nm.  The other interference filters allow us to monitor respectively the \(v' \! =\!  0\) to \(v \! =\!  1\) and \(v' \! =\!  0\) to \(v \! =\!  2\) decays at 275.2 nm and 278.8 nm. In order of increasing wavelength these three, 2 inch diameter filters (Andover Corporation) have center wavelengths of 271.6 nm, 275.3 nm, and 279.3 nm, and FWHM bandwidths of 10 nm, 1.5 nm and 2.0 nm.  We tune the laser to achieve the largest possible fluorescence signal.  This occurs at 271.7 nm in a region of the rotational spectrum where many of the low rotational number Q-branch transitions are simultaneously excited by our broad-band laser.  We refer to this region as the Q-branch bandhead.  We repeatedly compare the numbers of photon detections with each of our three filters, with the excitation laser tuned both on and off resonance.  The numbers are corrected for leakage of the strong transition at 271.7 nm through the other filters.   Taking into account the transmissions of the filters at the various wavelengths and the finite angular spread of the fluorescence passing through the filters, we infer the branching ratios reported in Table~\ref{tab02}.
The agreement with the predicted values is excellent. Figure~\ref{fig01}
summarizes our calculated and experimental results for the branching fractions for the decay of the \(B^{3}\Pi_{1}(v' \! =\!  0)\) state.  Nearly 99\% of the excited state population should return to the \(v \! =\!  0\) level and about 1\% will return to the \(v \! =\!  2\) level.  The decays to \(v \! =\!  1, 3, 4, 5\) and 6 or greater should all be less than a few hundredths of a percent.

\begin{figure}
\centering
\includegraphics[width=3.25in]{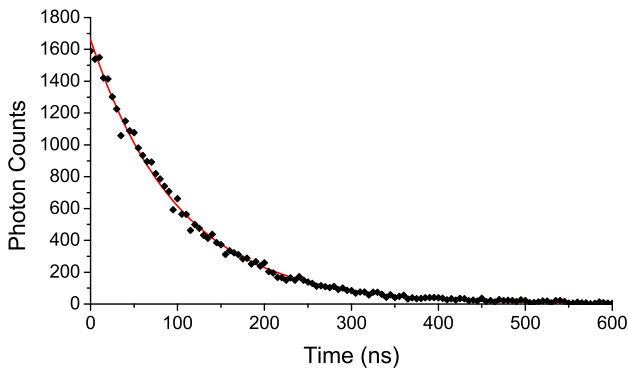}
\caption{\small (Color online.)  A typical lifetime measurement on the Q-branch bandhead. The points are the number of photon counts recorded in successive 5 ns bins.  The initial time in the fit and plot is chosen to be 110 ns after the laser pulse.  The data has been accumulated over 32,767 laser pulses and corrected for an offset and an electronics background.  The solid line is a simple exponential fit with a 101 ns decay time.}
\label{fig02}
\end{figure}

The \textit{B} state lifetime also plays a critical role in assessing possible laser cooling schemes.  To measure the lifetime, we feed our photomultiplier output into a multi-channel scalar (Stanford Research SR430).  The instrument sorts the emitted photon counts into sequential 5 ns bins.  We fit this data with three parameters describing an exponential decay plus a constant background.  Care is taken to avoid photon pile-up problems at the beginning of the decay. We measure the lifetime on the Q-branch bandhead as well as on the isolated rotational lines R23, R33, R43 and P61.  Each measurement is done using two different phototube/amplifier assemblies.  As a further check, additional lifetime measurements are made by simply averaging the decay fluorescence on a digital 300 MHz oscilloscope for the Q-branch bandhead and the lines R13, R23 and R33.  The statistical uncertainty in the lifetime associated with any individual fit (Figure~\ref{fig02})
is typically less than a few percent.  We observe some variations in the fit lifetimes with the detection and analysis system used and the selection of the time interval chosen for the fit.  The majority of these variations appear to be associated with our inability to completely remove an electronics background associated with the firing of the laser Q switch.  We have chosen to quote an uncertainty in our measured lifetime that encompasses all of the observed variations.  Combining the results from all of our measurements we conclude that the excited state lifetime is 99 (9) ns.  To the best of our knowledge, this is the first lifetime measurement of the B state of TlF.

Next we consider the implications of our results for laser cooling and slowing of a TlF beam, and what this would mean for future experiments using TlF in tests of fundamental symmetries.  We discuss two possible scenarios: one which simply uses an improved molecular beam, and another in which the TlF molecules could be slowed and trapped for much longer coherence time.  First, consider the possibility of using lasers to transversely cool and collimate a beam of TlF.  Our results suggest that with a cooling laser tuned to the Q1 line \(X(v \! =\!  0, J \! =\!  1, F_{1},F)\) to \(B(v' \! =\!  0, J' \! =\!  1, F'_{1} \! =\!  \frac{1}{2}, F')\) transition at 271.7 nm and a single repump laser, tuned to the analogous HFS component of the Q1 line but on the \(X(v \! =\!  2)\) to \(B(v' \! =\!  0)\) transition at 278.8 nm, one should be able to achieve several thousand cycles on the cooling transition.  This will be sufficient for transverse cooling and collimation of a cryogenic TlF beam \cite{ref14}.  Such a beam would have significantly increased occupation of the low rotational and vibrational levels of the molecule, as well as significantly lower forward velocity, compared to previous work with TlF beams.  Based on typical brightness, velocity distribution, and internal temperature achieved with this type of molecular beam source \cite{ref14}, plus a typical transverse velocity capture range (corresponding to a transverse temperature $\sim 5$ mK) and final transverse temperature ($\lesssim 300 \mu$K) of laser cooling \cite{ref01}, we estimate crudely that an EDM experiment conducted in such a collimated cryogenic beam might achieve approximately two orders of magnitude improvement over the present TlF limit.  This would yield limits on the proton EDM and Schiff moments comparable to those that have been achieved in the Hg EDM experiment \cite{ref15}.

Dramatic improvement in interaction time could be achieved if the beam can be stopped and trapped.  A TlF cryogenic beam would likely have a mean velocity of about \(\bar{v} \approx 150\)~m/s \cite{ref14}.  Approximately \(N = m\bar{v}\lambda/h \approx 23,000\) photon absorption/emission cycles are needed to stop these molecules.  To achieve this large number of cycles will likely require additional repump lasers to keep the molecules from accumulating in the \(X(v \! =\!  1, 3, 4 \mbox{~ and/or~ } 5)\) levels.  These levels can be repumped either through excitations to the \(B(v' \! =\!  0 \mbox{~ or~ } 2)\) levels or the \(A^{3}\Pi_{0^+} (v' \! =\!  0)\) state (see Table~\ref{tab03}) \cite{ref16}. Hence, with a few additional lasers it is likely that the beam could be stopped using the radiative force from photon scattering.  Further measurements of the vibrational branching fractions  for decays from \(B(v' \! =\!  0)\) to \(X(v \! =\!  1, 3, 4 \mbox{~ and/or~ } 5)\), at a level of sensitivity beyond what we were able to achieve, will be needed to evaluate exactly what lasers will be needed for such a task.
\begin{table*}
\begin{tabular}{|c||c|c|c|c|c|c|}\hline\hline
\(v'\backslash v\) & 0 & 1 & 2 & 3 & 4 & 5\\ \hline\hline
0 & 0.89(1) & 0.087(1) & 0.018(1) & 0.0024(3) & 0.0003(1) & 0.00004(1)\\ \hline
1 & 0.11(1) & 0.75(2) & 0.10(1) & 0.036(1) & 0.005(1) & 0.0010(2)\\ \hline
2 & 0.0003(1) & 0.15(1) & 0.71(2) & 0.075(6) & 0.051(3) & 0.007(1)\\ \hline
\end{tabular}
\caption{\small The Franck-Condon Factors for the \(X^{1}\Sigma^{+} - A^{3}\Pi_{0^+}\) transition of TlF calculated using Morse potentials derived from the spectroscopic parameters of Ref.~\cite{ref16}.  The primed (unprimed) vibrational quantum numbers refer to the excited (ground) electronic level.  The uncertainties in the last digits listed in parentheses reflect the uncertainties in the spectroscopic data used to determine the Morse potential and do not reflect any limitations associated with the Morse potential model.\label{tab03}}
\end{table*}

With knowledge of the excited state lifetime, we can estimate the length of the apparatus required to carry out radiative force slowing.  For a two level system, optimized spontaneous cooling allows one photon cycle to be completed in every two lifetimes of the excited state.  In our \(X(J \! =\!  1)\) level of the ground state, the existence of dark Zeeman states will cause the photon cycling rate to decrease by about a factor of three \cite{ref02}.  (The remixing of these dark states into the optical cycle can be accomplished by rapid switching of the slowing lasers' polarization \cite{ref17} or by resonant microwave transfer via the \(X(J\! =\! 0)\) level \cite{ref02}.) We estimate the length of the slowing region then would need to be about \(L \approx 3 \left(\frac{\bar{v}}{2}\right) 2\tau N \approx 1\)~m, which is a reasonable length for a molecular beam apparatus.

Alternatively, the additional lasers for repumping the \(X(v \! =\!  1, 3, 4 \mbox{~ and/or~ } 5)\) levels might be avoidable if other techniques for beam deceleration such as stimulated slowing \cite{ref18} or Stark deceleration \cite{ref19} were employed to slow the beam.  However, each of these would certainly introduce additional complications.  Overall, it is difficult to estimate the possible improvement of sensitivity to T-odd effects with an approach like this, since it will depend on details such as efficiency of trap loading, lifetime in the trap, etc.  Nominally, to achieve a factor of $\sim\! 100$ improvement in sensitivity would require trapping \(\sim\! 10^{7}\) molecules with lifetime $\sim\! 1$ sec.  This is similar to what has been achieved using Stark deceleration to load a storage ring of molecules, even without the additional advantage of laser cooling (albeit using species with parameters chosen to optimize the efficiency of slowing and trapping) \cite{ref20}.  Hence, we believe that such an approach may be viable and deserves further investigation.

In summary, the \(X^{1}\Sigma^{+} - B^{3}\Pi_{1}\) transition looks quite promising for the laser cooling of TlF.  While the simultaneous operation of several narrowband ultraviolet lasers will be costly, present laser technology can provide adequate power at the required excitation frequencies.  If the cooling is successful, this should open up the possibility of a new generation of fundamental symmetries experiments in TlF with significantly improved sensitivity.

We wish to thank Dr. D. Krause, Jr., R. Cann and N. Page for technical assistance and T. Shimasaki for calculations of RKR FCFs.  This work was supported by funds from Amherst College; the National Science Foundation under Grant Nos. PHY-0855465 and PHY-1068575; ARO; and AFOSR-MURI.

\end{document}